\documentclass[twocolumn,pra,showpacs,superscriptaddress]{revtex4}

\usepackage{graphicx}
\usepackage{dcolumn}
\usepackage{bm}

\begin{document}

\title{Superfluid and magnetic states of an ultracold Bose gas with synthetic three-dimensional spin-orbit coupling in
an optical lattice}


\author{Dan-Wei Zhang}
\affiliation{Laboratory of Quantum Information Technology and SPTE,
South China Normal University, Guangzhou 510006, China}
\affiliation{Department of Physics and Center of Theoretical and
Computational Physics, The University of Hong Kong, Pokfulam Road,
Hong Kong, China}

\author{Ji-Pei Chen}
\affiliation{Laboratory of Solid State Microstructures and
Department of Physics, Nanjing University, Nanjing 210093, China}

\author{Chuan-Jia Shan}
\affiliation{Laboratory of Quantum Information Technology and SPTE,
South China Normal University, Guangzhou 510006, China}
\affiliation{College of Physics and Electronic Science, Hubei Normal
University, Huangshi 435002, China}

\author{Z. D. Wang}
\email{zwang@hku.hk}\affiliation{Department of Physics and Center of
Theoretical and Computational Physics, The University of Hong Kong,
Pokfulam Road, Hong Kong, China}

\author{Shi-Liang Zhu}
\email{slzhu@scnu.edu.cn} \affiliation{Laboratory of Quantum
Information Technology and SPTE, South China Normal University,
Guangzhou 510006, China}
\affiliation{Laboratory of Solid State
Microstructures and Department of Physics, Nanjing University,
Nanjing 210093, China}

\begin{abstract}
We study 
ultracold bosonic atoms with the synthetic three-dimensional
spin-orbit (SO) coupling in a cubic optical lattice. In the
superfluidity phase, the lowest energy band exhibits one, two or
four pairs of degenerate single-particle ground states depending on
the SO-coupling strengths, which can give rise to the condensate
states with spin-stripes for the weak atomic interactions. In the
deep Mott-insulator regime, the effective spin Hamiltonian of the
system combines three-dimensional Heisenberg exchange interactions,
anisotropy interactions, and Dzyaloshinskii-Moriya interactions.
Based on Monte Carlo simulations, we numerically demonstrate that
the resulting Hamiltonian with an additional Zeeman field has a rich
phase diagram with spiral, stripe, vortex crystal, and especially
Skyrmion crystal spin-textures in each $xy$-plane layer. The
obtained Skyrmion crystals can be tunable with square and hexagonal
symmetries in a columnar manner along the $z$ axis, and moreover are
stable against the inter-layer spin-spin interactions in a large
parameter region.
\end{abstract}

\date{\today}
\maketitle

\section{introduction}

Spin-orbit (SO) coupling plays an important role in condensed matter
physics, especially in the newly discovered quantum spin Hall
effects and the related topological orders \cite{qi2010,kane2010}.
Recent theoretical proposals \cite{dalibard2011,Zhu2006} and
experimental realization \cite{lin2011,chen2012,wang2012,cheuk2012}
of non-Abelian gauge fields in ultracold atoms with the optical
dressing technique open another door to explore SO coupling in
controllable systems. The bulk gases of weakly interacting bosons
with the synthetic two-dimensional (2D) Rashba SO coupling in
homogenous cases and in trapping potentials have been widely studied
and theoretically shown to exhibit exotic many-body ground states
\cite{Zhai2012}, some of which have no direct analog in solid-state
systems \cite{wang2010,wu2008,santos2011,hu2012}. For example, an
SO-coupled spinor condensate will spontaneously develop a plane wave
phase or spin-stripe structure depending on the weak interaction
energy \cite{wang2010,wu2008}; and in the presence of strong
trapping potentials, it will exhibit half-quantum vortex states and
Skyrmion patterns \cite{wu2008,santos2011,hu2012}.

Recently, physics of a SO-coupled bosonic gas loaded in a 2D optical
lattice (OL) has attracted considerable interest
\cite{larson2010,Lewenstein2011,cole2012,radic2012,cai2012,gong2012,mandal2012,martikainen2012,liu2012,wong2012}.
This system can be described by an extended two-component
Bose-Hubbard (BH) model \cite{fisher1989,Jaksch1998,Greiner2002}, in
which the SO coupling can significantly affect the quantum phase
transition from a superfluid to a Mott insulator (MI)
\cite{Lewenstein2011,cole2012,mandal2012}. More interestingly, the
effective spin Hamiltonian of the system in the deep MI regime
contains the so-called Dzyaloshinskii-Moriya (DM) interaction term
\cite{DM}, which comes from the SO coupling and may lead to some
novel magnetic phases \cite{cole2012,radic2012,cai2012,gong2012},
such as Skyrmion crystals \cite{cole2012}. However, the
corresponding three-dimensional (3D)  system is yet to be explored.

On the other hand, the 3D analog of SO coupling in cold atoms has
been proposed to be experimentally realized by using optical
dressing schemes \cite{anderson2012,liy2012} and by exploiting
laser-assisted tunneling \cite{jaksch2003} in OLs
\cite{yang2010,bermudez2010}. The 3D SO-couplings are less explored
in contrast to the standard 2D Rashba and Dresshauls ones in the
context of condensed matter physics \cite{anderson2012}, but is now
attracting more and more interests
\cite{kane2010,balents2011,han2011,shenoy2011} for investigating 3D
topological insulators \cite{kane2010}, topological superfluidity
\cite{han2011}, Weyl semi-metals \cite{balents2011}, and
spherical-SO-coupling-induced Bardeen-Cooper-Schreiffer to
Bose-Einstein condensate (BEC) crossover \cite{shenoy2011}. Very
recently, several pieces of theoretical work study the ground state
of a weakly-interacting two-component BEC with the 3D SO-coupling in
the continuum \cite{kawakami2012,liyi2012,bmanderson2012}. It is
shown that the density distribution of the ground state BEC can also
exhibit the interesting Skyrmion structure, which is moreover a 3D
counterpart characterized by a 3D topological winding number
\cite{kawakami2012,liyi2012}. Some other schemes have also been
proposed to create Skyrmions in multicomponent BECs in the absence
of synthetic SO-couplings and OLs \cite{khawaja2001}. So it would be
worthwhile to search the stable Skyrmion (crystals) in an OL system
with 3D SO-coupled bosons.

In this paper, we investigate ultracold bosons with the synthetic 3D
SO coupling in a cubic OL. We first look into the weakly interacting
superfluidity case. In this case, the lowest energy band exhibits
one, two or four pairs of degenerate single-particle ground states
related to the SO coupling, and each pair contains opposite wave
vectors with values depending on the SO-coupling strengths. This can
give rise to the condensate states with spin-stripes for the weak
atomic interactions. We then focus on the deep MI regime with one
atom per lattice and derive the effective spin-spin interaction
Hamiltonian of the system. The spin Hamiltonian is a combination of
three-dimensional Heisenberg exchange interactions, anisotropy
interactions and DM interactions. Based on Monte Carlo (MC)
simulations, we numerically demonstrate that the resulting
Hamiltonian with an additional Zeeman field has a rich classical
phase diagram with spiral, stripe, vortex crystal, and especially
Skyrmion crystal spin-textures in each $xy$-plane layer. We find
that the obtained Skyrmion crystals can be square or hexagonal
symmetries with experimentally tunable parameters by varying
laser-atom interactions. Moreover, the Skyrmion crystals in a
columnar manner of extending along $z$ axis are stable against the
inter-layer spin interactions within a large parameter region. This
cold atom system with high controllability in the effective spin
interactions may provide an ideal platform to further study exotic
quantum spin models and find new phases of matter.

The paper is organized as follows. The next section (Sec. II)
introduces an extended BH model which describes cold bosons with the
synthetic 3D SO coupling in a cubic OL. In Sec. III, we briefly
analyze the single-particle energy band and the properties of the
weakly interacting superfluidity phase. In Sec. IV, we derive the
spin Hamiltonian of the system in the MI regime, and present its
rich classical phase diagram with interesting spin configurations.
In particular, we study the profiles and stability of the Skyrmion
crystals in the system. A brief discussion and short conclusion are
finally given in Sec. V.

\section{model}

Let us consider an atomic gas of pseudospin-1/2 ultracold bosons
loaded into a 3D cubic optical lattice with the synthetic SO
coupling. The single-particle Hamiltonian of the system is written
as
\begin{equation}
\label{SinHam} \hat{H}_0=\frac{\hat{{\textbf
p}}^2}{2m}+\kappa_x\sigma_x\hat{p}_x+\kappa_y\sigma_y\hat{p}_y+\kappa_z\sigma_z\hat{p}_z+V(x,y,z),
\end{equation}
where $m$ is the atomic mass, $\hat{{\textbf p}}$ is the momentum
operator, $\kappa_{\eta}$ with $\eta=x,y,z$ is respectively the
strength of SO coupling along the $\eta$ axis, and
$\sigma_{x,y,z}$ are the three Pauli matrices. Here the cubic
optical lattice $V(x,y,z)=\sum_{\eta}V_{\eta}\sin^2(k_0\eta)$ is
formed by three standing-wave laser beams with the same wave
number $k_0$. Thus the lattice spacing is $a=\pi/k_0$. Hamiltonian
(\ref{SinHam}) can be rewritten as $\hat{H}_0=(\hat{{\textbf
p}}-{\textbf A})^2/2m+V$ up to a constant, where the non-Abelian
gauge potential ${\textbf
A}=-m(\kappa_x\sigma_x,\kappa_y\sigma_y,\kappa_z\sigma_z)$. This
is corresponding to a 3D SO coupling.

We consider the system in the tight-binding regime, which is
reachable in realistic experiments \cite{Jaksch1998,Greiner2002}.
Under this condition, in the presence of such a non-Abelian gauge
field, the bosons can be described by an extended single-band BH
Hamiltonian in terms of Peierls substitution
\cite{cole2012,radic2012}:
\begin{equation}
\label{3DHam} \mathcal{H}=-\sum_{{\bf
i},\hat{\eta}}\left[t_{\eta}\hat{a}^{\dag}_{{\bf
i},\sigma}\mathcal{R}_{\eta}^{\sigma\sigma'}\hat{a}_{{\bf
i}+\hat{\eta},\sigma'}+\text{H.c.}\right]+V_{\text{int}},
\end{equation}
where $\hat{a}^{\dag}_{{\bf i},\sigma}$ ($\hat{a}_{{\bf i},\sigma}$)
creates (annihilates) a spin-$\sigma$ ($\sigma=\uparrow,\downarrow$)
boson at the site ${\bf i}$. The first term in Hamiltonian
(\ref{3DHam}) describes atomic hopping between neighbor lattice
sites, with $t_{\eta}$ representing the overall hopping amplitude in
the absence of the synthetic SO coupling. The $2\times2$ matrix
$\mathcal{R}_{\eta}\equiv \exp(-\frac{i}{\hbar}A_{\eta}a)$ is the
Peierls substitution along direction $\hat{\eta}$ with respect to
the gauge potential. We rewrite
$\mathcal{R}_{\eta}=\exp(i\theta_{\eta}\sigma_{\eta})=\cos\theta_{\eta}\mathbf{1}+i\sigma_{\eta}\sin\theta_{\eta}$
with dimensionless SO-coupling strength $\theta_{\eta}=\pi
m\kappa_{\eta}/\hbar k_0$. The diagonal and off-diagonal terms in
the matrix respectively refer to the spin-conserving hopping and
spin-flip hopping due to the SO coupling in the $xy$ plane. The
second term in Hamiltonian (\ref{3DHam}) denotes the atomic
repulsive interactions, which is given by
\begin{equation}
\label{Vint}
V_{\text{int}}=\frac{1}{2}\sum_{{\bf
i},\sigma\sigma'}U_{\sigma\sigma'}\hat{a}^{\dag}_{{\bf
i},\sigma}\hat{a}^{\dag}_{{\bf i},\sigma'}\hat{a}_{{\bf
i},\sigma'}\hat{a}_{{\bf i},\sigma},
\end{equation}
where $U_{\sigma\sigma'}$ is interaction strength between spins
$\sigma$ and $\sigma'$. The atomic interactions are almost
spin-independent in experiments in the absence of Feshbach
resonances \cite{lin2011,chen2012}, and hence we assume
$U_{\uparrow\uparrow}=U_{\downarrow\downarrow}\equiv U$ and
$U_{\uparrow\downarrow}=U_{\downarrow\uparrow}=\alpha U$ with
$\alpha\approx1$. In fact, the slight difference between the
intraspecies and interspecies interaction strengths will help to
select the degenerate many-body ground states of the system in the
weak interacting superfluidity phase.

This extended 3D BH model also exhibits the superfluidity and the MI
phases for the weak and strong atomic interactions compared with the
hopping energy \cite{Jaksch1998,Greiner2002}, respectively. The
quantum phase transitions between them are affected by the SO
coupling in a similar manner as that in the 2D cases
\cite{Lewenstein2011,cole2012,mandal2012}. Thus we just consider the
system in the two interaction limits in this work, and focus on the
effects of synthetic 3D SO-coupling in the superfluidity and the MI
phases.

\section{superfluidity states}

In this section, we consider the weakly-interacting superfluidity
phase. In this regime, the hopping term dominates in Hamiltonian
(\ref{3DHam}). We first look into the hopping Hamiltonian
$\mathcal{H}_T=\mathcal{H}-V_{\text{int}}$ to obtain the energy band
of the system, and then briefly discuss the effects of weak atomic
interactions. The corresponding Hamiltonian in the momentum space
can be written as
\begin{equation}
\mathcal{H}_{\bf k}=\sum_{\bf k}\left(\hat{a}^{\dag}_{{\bf
k},\uparrow}~\hat{a}^{\dag}_{{\bf k},\downarrow}\right)~\hat{H}_{\bf
k}~\left(
                                        \begin{array}{c}
                                          \hat{a}_{{\bf
k},\uparrow} \\
                                          \hat{a}_{{\bf
k},\downarrow} \\
                                        \end{array}
                                      \right).
\end{equation}
By using spatial Fourier transformations on $\mathcal{H}_T$, we
can obtain $\hat{H}_{\bf k}={H}_{kx}+{H}_{ky}+{H}_{kz}$ with
$H_{k\eta}=-2t_{\eta}\cos\theta_{\eta}\cos(k_{\eta} a)\mathbf{1}
+2t_{\eta}\sin\theta_{\eta}\sin(k_{\eta}a)\sigma_{\eta}$. 
Diagonalizing $\hat{H}_{\bf k}$ yields the energy structure of the
system in the vanishing interaction limit:
\begin{eqnarray}
\label{Ek} E_{{\bf
k}}^{\pm}=-2\sum_{\eta}t_{\eta}\cos\theta_{\eta}\cos(k_{\eta}a)~~~~~~\nonumber\\
\pm2\sqrt{\sum_{\eta}t^2_{\eta}\sin^2\theta_{\eta}\sin^2(k_{\eta}a)}.
\end{eqnarray}

The lowest energy states in the resulting lower Bloch band $E_{\bf
k}^-$ present candidates for the many-body ground-state of the
bosonic gas with weak interatomic interactions. The Bloch momentum
of these states denoted by ${\textbf k}_0\equiv(k_0^x,k_0^y,k_0^z)$
can be directly obtained by solving the equation $\partial_{\bf
k}\left(E_{\bf k}^-\right)|_{{\bf k}={\bf k}_0}=0$ for minimizing
$E_{\bf k}^-$ with the specific parameters $t_{\eta}$ and
$\theta_{\eta}$. We find that there are possibly one, two or four
pairs of degenerate minima in the lowest energy band for different
values of $t_{\eta}$ and $\theta_{\eta}$ in the case of
non-vanishing SO coupling. Each pair come as a time-reversed partner
with opposite wave vectors. For simplicity, we focus on the
isotropic tunneling with $t_x=t_y=t_z$, the degeneracy of the ground
states can be split into three cases by the configuration of the SO
coupling \cite{shenoy2011}. The first case is that the SO coupling
is anisotropic along all three dimensions. We then have a single
pair of degenerate minima. If it is anisotropic along two
dimensions, whether we have one pair, or two pairs, depending on the
configuration of the SO coupling is {\sl prolate} or {\sl oblate}
\cite{shenoy2011}. Finally, the case of maximum symmetry with fully
isotropic SO coupling gives four pairs of degenerate minima. For
example, when $\theta_{x}=\theta_{y}<\theta_{z}$ (the {\sl prolate}
case), two degenerate minima locate at ${\textbf k}_0=(0,0,\pm
\theta_z/a)$; when $\theta_{x}=\theta_{y}>\theta_{z}$ (the {\sl
oblate} case), four degenerate minima locate at $(\pm \xi_1/a,\pm
\xi_1/a,0)$ with $\tan \xi_1=\tan \theta_x/\sqrt{2}$; when
$\theta_{x}=\theta_{y}=\theta_{z}$, eight degenerate minima locate
at $(\pm \xi_2/a,\pm \xi_2/a,\pm \xi_2/a)$ with $\tan \xi_2=\tan
\theta_x/\sqrt{3}$. This is in sharp contrast to the continuum case
where the rotationally symmetric dispersion has an infinite ground
sates degeneracy forming an SO sphere \cite{anderson2012}. Owing to
the reduction of degeneracy, the Bose condensates with the synthetic
3D SO-coupling in an OL would be more robust against quantum
fluctuations than their bulk counterparts.

For the weakly interacting cases (i.e., $U\ll t_{\eta}$) within the
Gross-Pitaevskill (GP) approximation, the analysis of the ground
state (condensate) wave-fucntion in this system is in parallel to
those of the counterparts in 2D OLs \cite{cole2012}, as well as in
2D and 3D continuum cases \cite{wang2010,liyi2012}. Hence we just
present the conclusions here without detailed calculations. The
condensate wave-function can be written as a superposition of all
possible single-particle (plane-wave) wave-functions of the the
lowest Bloch states discussed above, and the corresponding
superposition coefficients are determined by minimizing the
mean-field GP interaction (i.e., density-density interaction) energy
\cite{wang2010,liyi2012}. The GP interaction energy can be divided
into the spin-independent and spin-dependent parts. The
spin-independent term yields the same Hartree-Fork energy for any
different selection of superposition coefficients, but the
spin-dependent one with respect to $\alpha$ selects the coefficients
for minimizing itself. For $\alpha<1$, only one of the lowest
degenerate single-particle state (whose numbers can be two, four or
eight) is occupied, and thus the condensate wave-function is a
plane-wave state with a finite momentum. On the other hand, for
$\alpha>1$, one of the paired degenerate states (whose numbers can
be one, two or four) are occupied with equal superposition
coefficients, giving rise to the condensate states with spin-stripe
density distribution \cite{wu2008,wang2010}. The structure of the
spin-stripe is dependent on the vector ${\textbf k}_0$ and hence is
tunable by the synthetic SO coupling. We note that these ground
states are still degenerate except the case of stripe states with
only two degenerate minima in the Bloch band. To further remove this
accidental degeneracy, one should consider quantum fluctuations
\cite{barnett2012}.

\section{spin model and magnetic states in Mott-insulator regime}

\subsection{Effective spin Hamiltonian}

In this section, we turn to consider the system in the MI phase. We
are interested in the MI regime with $U\gg t_{\eta}$ and nearly unit
atom per lattice site. In this case, the atoms are localized in
individual lattices and the nearest-neighbor hopping can be treated
as a perturbation, leading to an effective spin Hamiltonian
\cite{Auerbach}. To obtain the spin Hamiltonian of the system, one
can begin with a two-site problem \cite{kuklov2003}. In the zero
order, the system is described by the interaction Hamiltonian
(\ref{Vint}), and the ground state manifold for the two-site problem
with one atom in each site composes four degenerate zero-energy
states
$\{|\uparrow;\uparrow\rangle,|\uparrow;\downarrow\rangle,|\downarrow;\uparrow\rangle,|\downarrow;\downarrow\rangle\}$.
Here we have assumed uniform in-site energy and chosen it as the
energy base. The exchange of two atoms in different sites does not
require energy, and hence the single atom hopping should be
eliminated in the second order with respect to the ratio
$t_{\eta}/U$. In this progress, there are six excited states
$\{|\uparrow\downarrow;0\rangle,|0;\uparrow\downarrow\rangle,
|\uparrow\uparrow;0\rangle,|0;\uparrow\uparrow\rangle,|\downarrow\downarrow;0\rangle,|0;\downarrow\downarrow\rangle\}$
with an energy $U$. The hopping perturbation described by
$\mathcal{H}_{\text T}$ couples the ground-state manifold and the
excited-state one. The resulting effective Hamiltonian up to the
second order of perturbation reads \cite{Auerbach,kuklov2003}
\begin{equation}
\label{EffHam}
(H_{\text{eff}})_{\beta\nu}=-\sum_{\gamma}\frac{(H_{\text
T})_{\beta\gamma}(H_{\text
T})_{\gamma\nu}}{E_{\gamma}-(E_{\beta}+E_{\nu})/2},
\end{equation}
where $\beta$ and $\nu$ label the four states in the ground-states
manifold and $\gamma$ labels the six excited ones.

After obtaining the two-site effective Hamiltonian from Eq.
(\ref{EffHam}), it is straightforward to extend it to the lattice
counterpart by introducing nearest neighbor hopping in the whole
lattice. It is convenient to write the lattice effective
Hamiltonian in terms of isospin operators $\vec{S}_{\bf i}=(S_{\bf
i}^x,S_{\bf i}^y,S_{\bf i}^z)$ with $S_{\bf
i}^x=\frac{1}{2}(\hat{a}^{\dag}_{{\bf i},\uparrow}\hat{a}_{{\bf
i},\downarrow}+\hat{a}^{\dag}_{{\bf i},\downarrow}\hat{a}_{{\bf
i},\uparrow})$, $S_{\bf i}^y=-\frac{i}{2}(\hat{a}^{\dag}_{{\bf
i},\uparrow}\hat{a}_{{\bf i},\downarrow}-\hat{a}^{\dag}_{{\bf
i},\downarrow}\hat{a}_{{\bf i},\uparrow})$, and $S_{\bf
i}^z=\frac{1}{2}(\hat{a}^{\dag}_{{\bf i},\uparrow}\hat{a}_{{\bf
i},\uparrow}-\hat{a}^{\dag}_{{\bf i},\downarrow}\hat{a}_{{\bf
i},\downarrow})$. The resulting spin Hamiltonian of this 3D system
in in the deep MI region is then given by
\begin{equation}
\begin{array}{ll}
\label{3DSHam} \mathcal{H}_s = -\displaystyle\sum_{\bf
i}\vec{S}_{\bf i}\cdot\left(J_x\vec{S}_{{\bf
i}+\hat{x}}+J_y\vec{S}_{{\bf i}+\hat{y}}+J_z\vec{S}_{{\bf
i}+\hat{z}}\right)\\
~~~~~~~-\displaystyle\sum_{\bf i}\left(K_x S^x_{\bf i}S^x_{{\bf
i}+\hat{x}}+K_y S^y_{\bf i}S^y_{{\bf i}+\hat{y}}+K_z S^z_{\bf
i}S^z_{{\bf
i}+\hat{z}}\right)\\
~~~~~~~-\displaystyle\sum_{\bf i}\left(D_x\vec{S}_{\bf
i}\times\vec{S}_{{\bf i}+\hat{x}}\cdot \hat{x}+D_y\vec{S}_{\bf
i}\times\vec{S}_{{\bf
i}+\hat{y}}\cdot \hat{y}\right.\\
~~~~~~~~~~~~~~~~~+\left.D_z\vec{S}_{\bf i}\times\vec{S}_{{\bf
i}+\hat{z}}\cdot \hat{z}\right),
\end{array}
\end{equation}
where $J_{\eta}=\frac{4t_{\eta}^2}{U}\cos(2\theta_{\eta})$,
$K_{\eta}=\frac{8t_{\eta}^2}{U}\sin^2\theta_{\eta}$, and
$D_{\eta}=\frac{4t_{\eta}^2}{U}\sin(2\theta_{\eta})$ are spin
interaction strengths. Hamiltonian (\ref{3DSHam}) describes a
generally anisotropic 3D spin-spin interaction system. It combines
the Heisenberg exchange interaction as the first term, the
anisotropy interaction as the second term, and 3D DM spin
interaction as the last term. Note that all the spin interaction
strengths in these terms are dependent on the laser beams which
generate the OL and the SO coupling, and hence they are tunable in
experiments.

To proceed further, we introduce an effective Zeeman term to
Hamiltonian (\ref{3DSHam}), leading to the total spin Hamiltonian
\begin{equation}
\label{TotHam} \mathcal{H}^T_s = \mathcal{H}_s-h_z\sum_{\bf i}S_{\bf
i}^z.
\end{equation}
This Zeeman term can be easily achieved by applying an additional
external field to the pseudospin-1/2 atoms. For the pseudospin
states that are usually two atomic hyperfine states, the external
field can be simply a real magnetic field
\cite{kuklov2003,duan2003}. If the pseudospin states are dressed
states, one can use an appropriately designed laser field to
generate it \cite{zhu2011}. Thus $h_z$ is also a tunable parameter.
We assume that the strength of Zeeman field here $h_z\ll U$ but is
comparable with $t_{\eta}^2/U$. Under this condition, the parameters
in Hamiltonian (\ref{3DSHam}) are approximately unchanged.

\begin{figure}[tbph]
\includegraphics[width=0.8\linewidth]{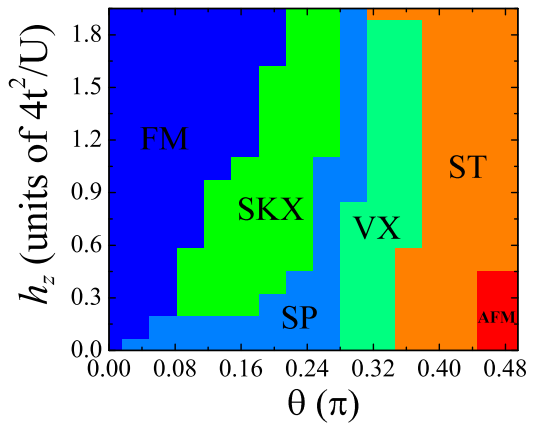}
\caption{(Color online) Classical phase diagram of the reduced 2D
layer version of the spin Hamiltonian $\mathcal{H}^T_s$ with
$t_z=0$, obtained from Monte Carlo simulations. This is related to
the bosons loaded in the OL in the deep MI regime. The abbreviations
are: FM: ferromagnetic phase; AFM: antiferromagnetic phase; SP:
spiral phase; ST: stripe phase; VX: vortex crystal phase; and SKX:
Skyrmion crystal phase. Their definitions are in the text. Same
typical spin configurations of unusual phases (SP, ST, SKX, and VX)
are shown in Fig. 2 and Fig. 3.} \label{fig:phase}
\end{figure}


\subsection{Numerical results from Monte Carlo simulations}

Below we explore the low-temperature phase diagrams and ground
states of the effective spin Hamiltonian $\mathcal{H}^T_s$ [see Eq.
(\ref{TotHam})], and we also intend to find stable Skyrmionic
spin-textures via MC simulations. In this classical approximation we
treat the spins $\vec{S}_{\bf i}$ as classical unit vectors and aim
to find the spin configurations $\{\vec{S}_{\bf i}\}$ for minimizing
the energy. We note that the classical MC simulation has been used
to explore the phase diagrams of several kinds of spin models, such
as the Heisenberg model with DM interactions in the context of
solids \cite{yi2009,roler2006,yu2010}. While this method may not be
used to determine the precise phase boundaries and to search for
some phases driven by quantum fluctuations, it can be an efficient
tool to determine different possible phases, especially when there
are no degeneracies on the classical level. For example, the
classical MC results in Ref. \cite{cole2012} is consistent with
those by the variational analysis in Ref. \cite{cai2012}. Thus one
can expect that the classification of ground states in our MC
simulations could generally survive in a variational approach, but a
full quantum treatment would be needed to have a deeper
understanding of the quantum spin model (\ref{TotHam}), which is
beyond the present work.


\begin{figure}[tbph]
\includegraphics[width=0.52\linewidth]{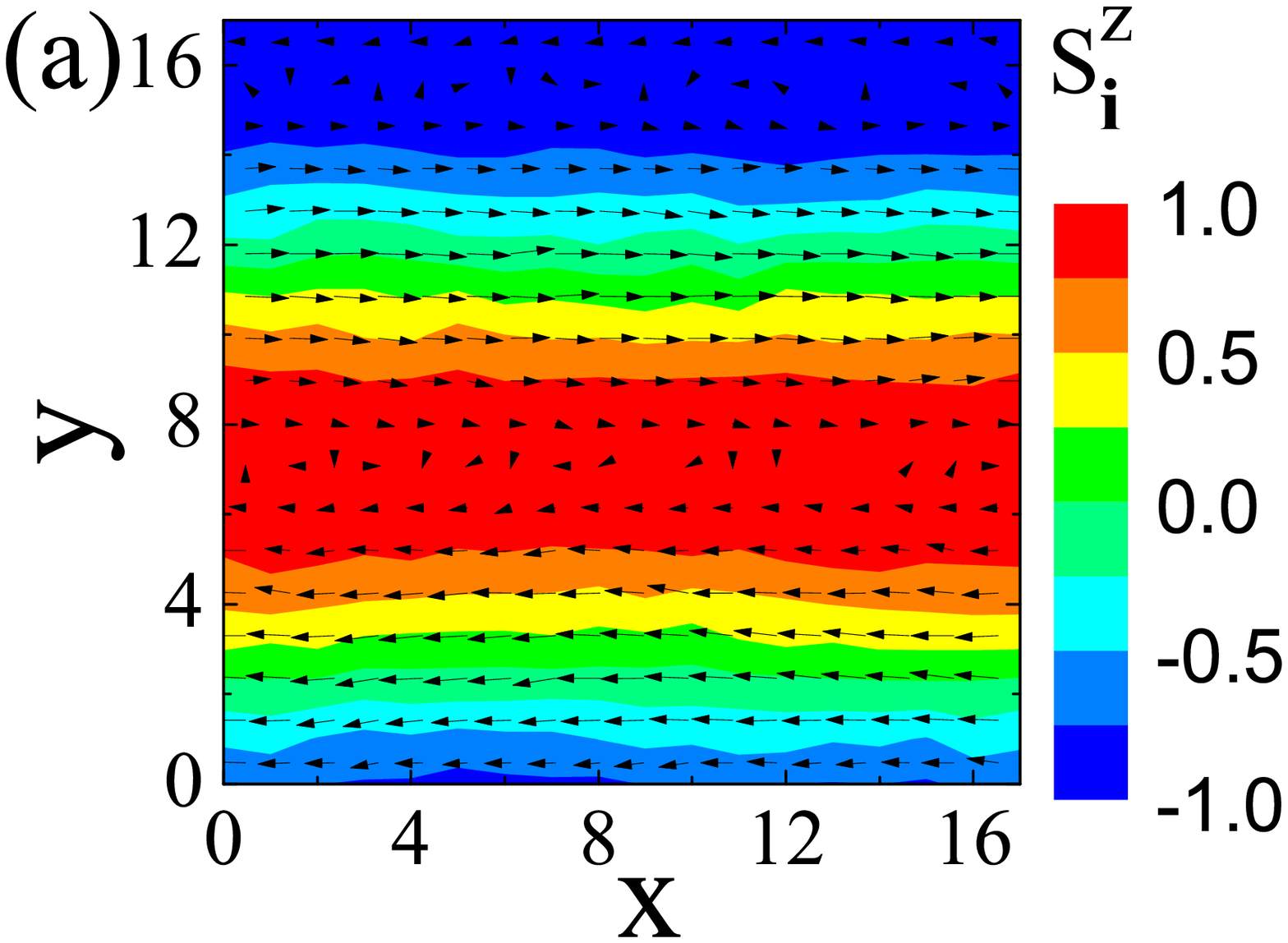}
\includegraphics[width=0.42\linewidth]{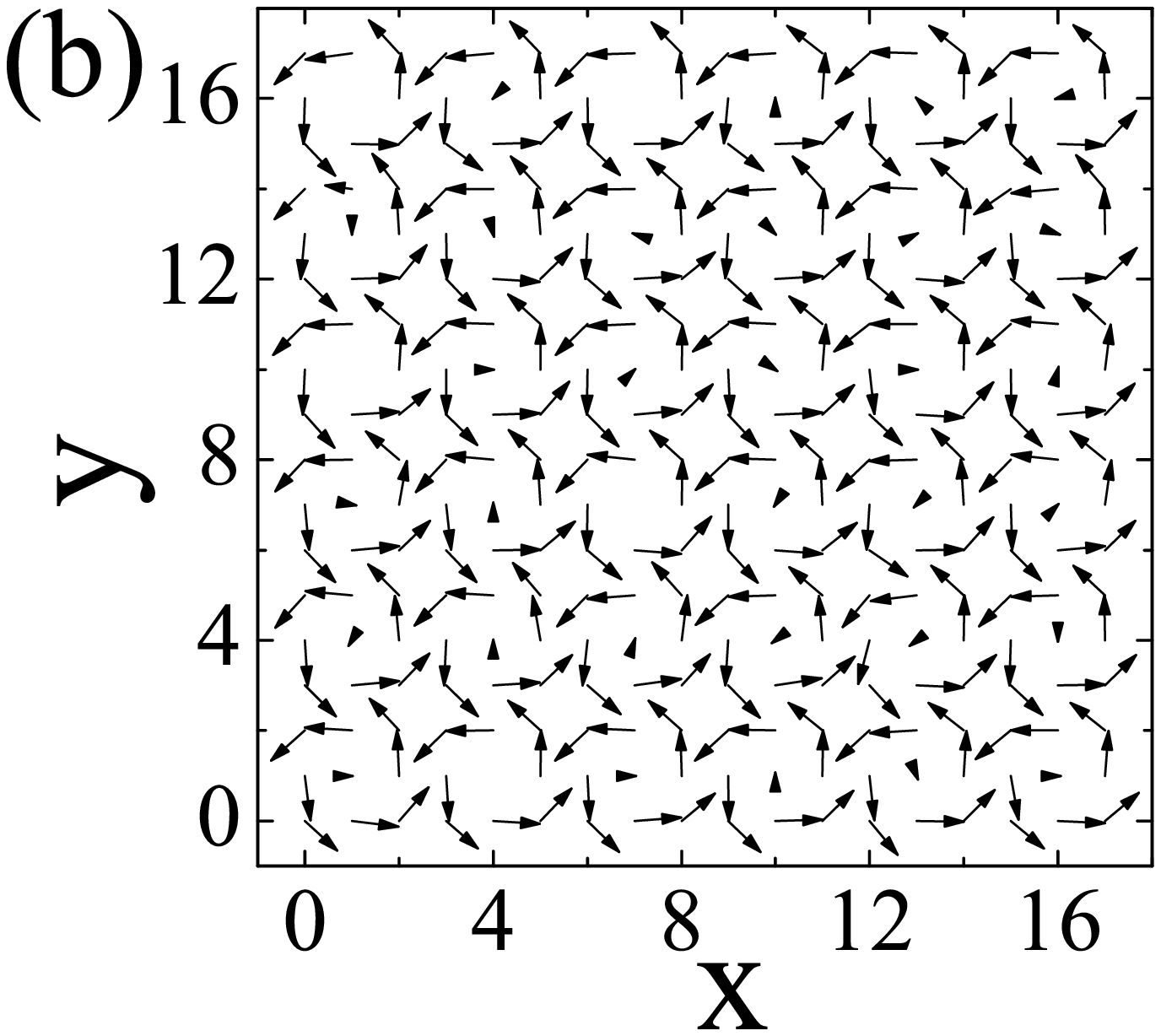}
\includegraphics[width=0.45\linewidth]{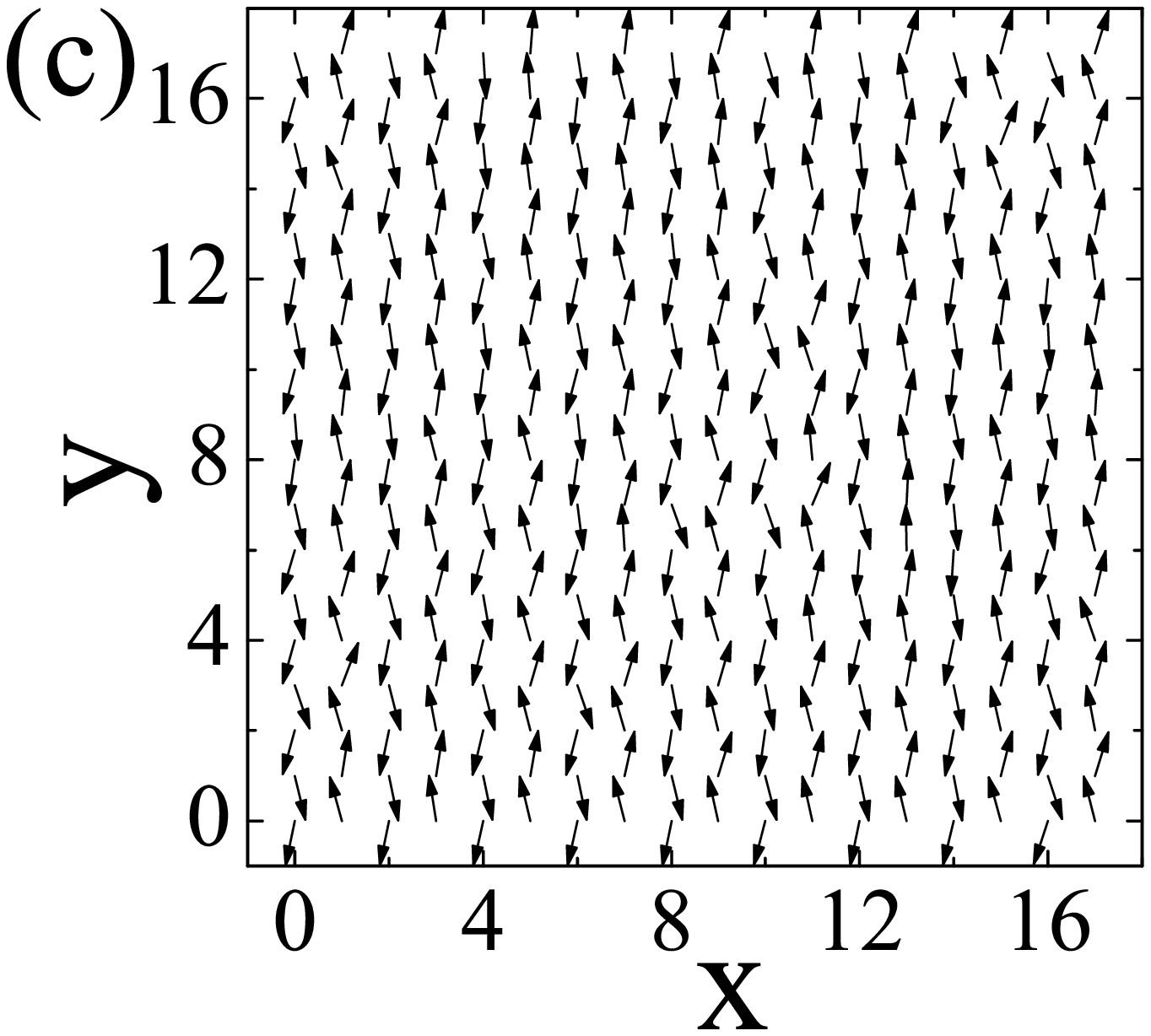}~~~
\includegraphics[width=0.45\linewidth]{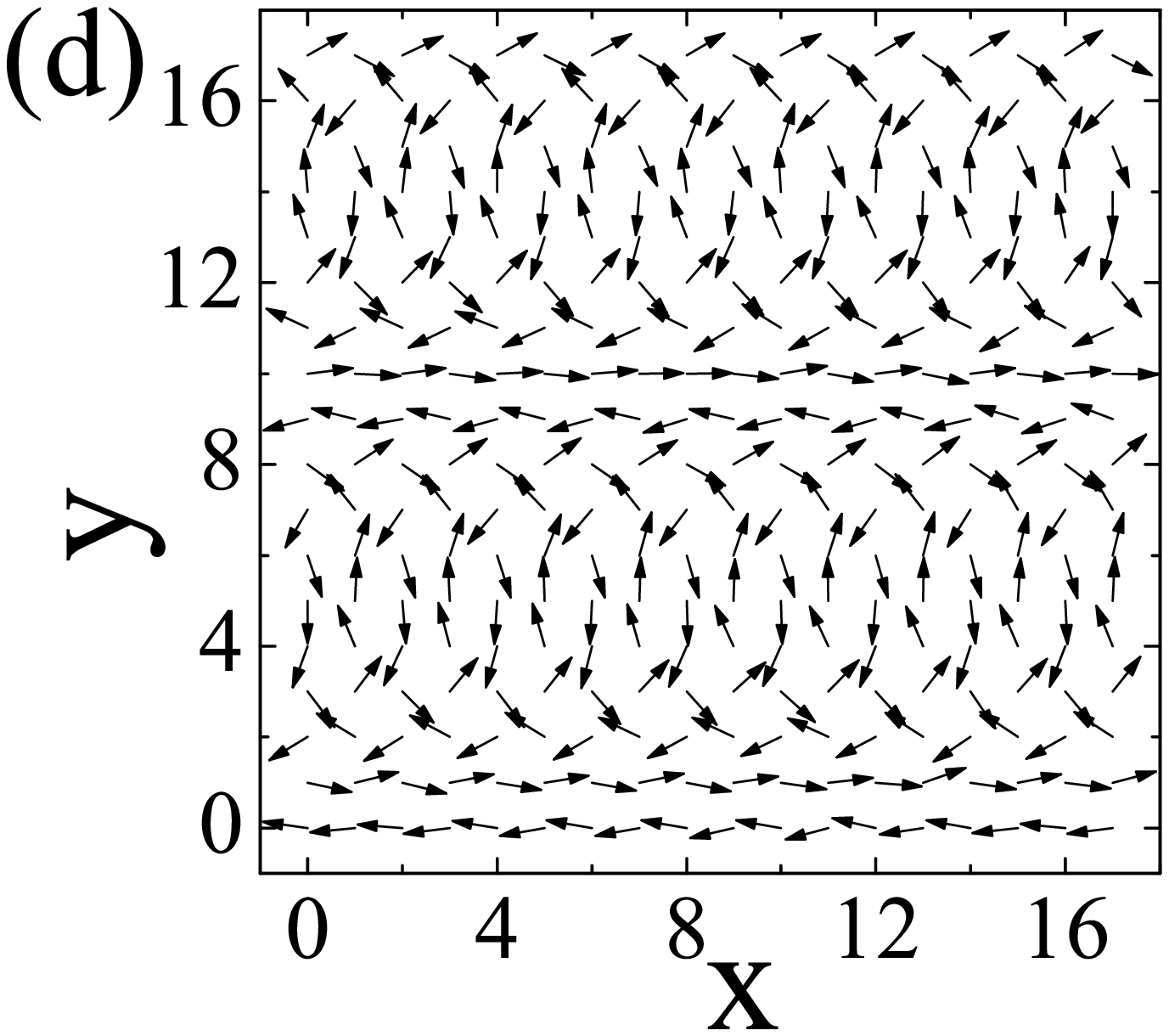}
\includegraphics[width=0.26\linewidth]{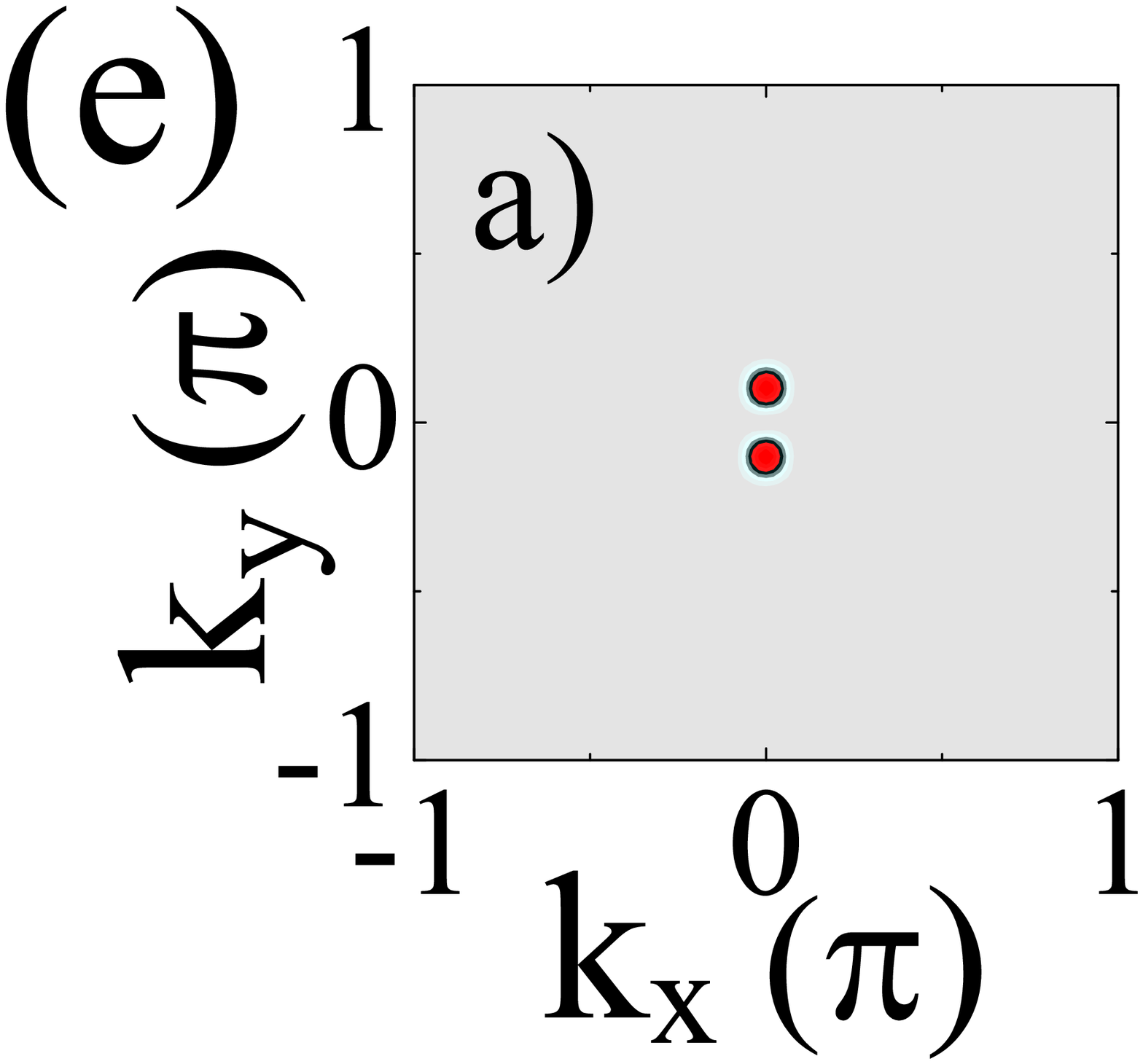}
\includegraphics[width=0.2\linewidth]{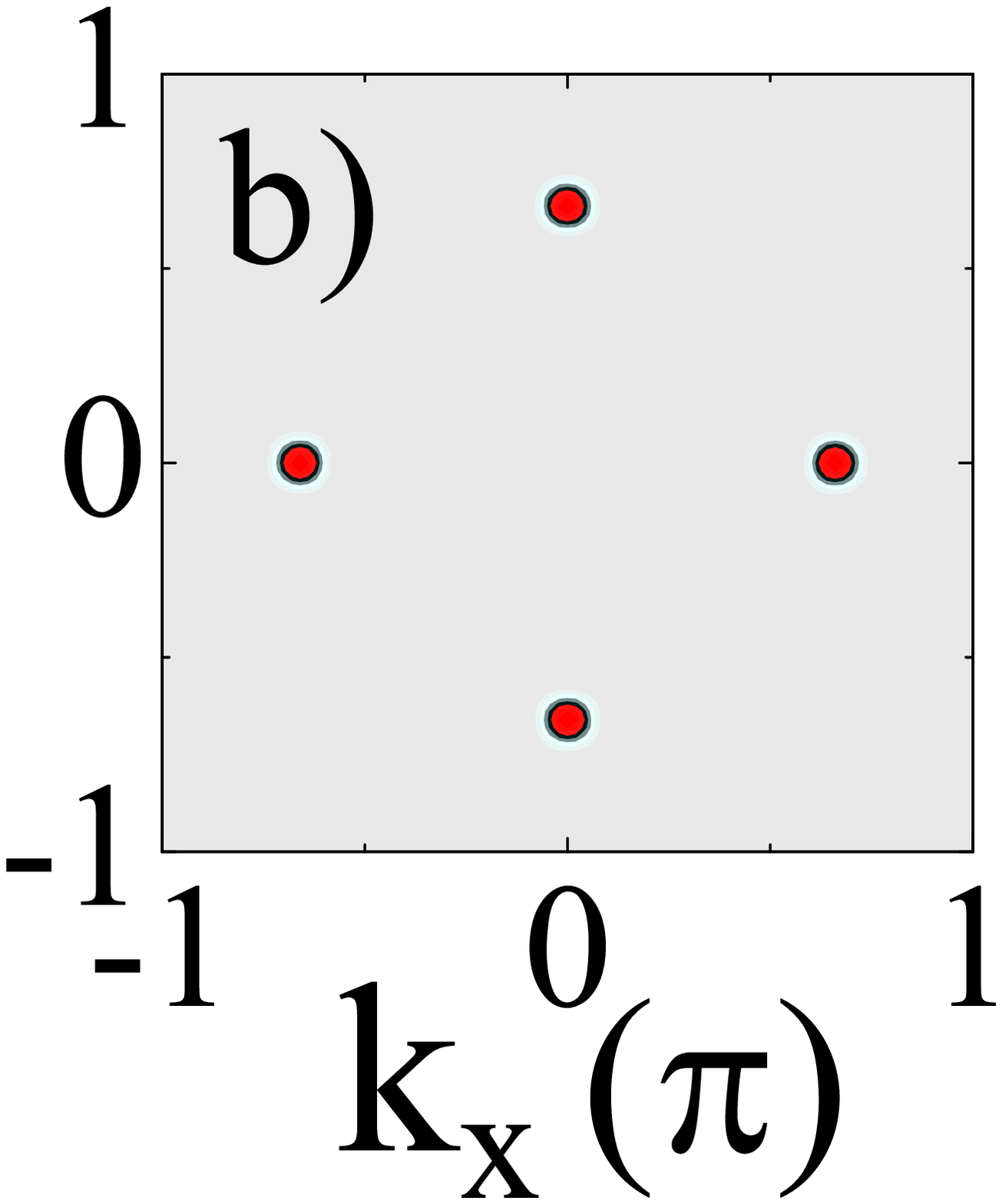}
\includegraphics[width=0.2\linewidth]{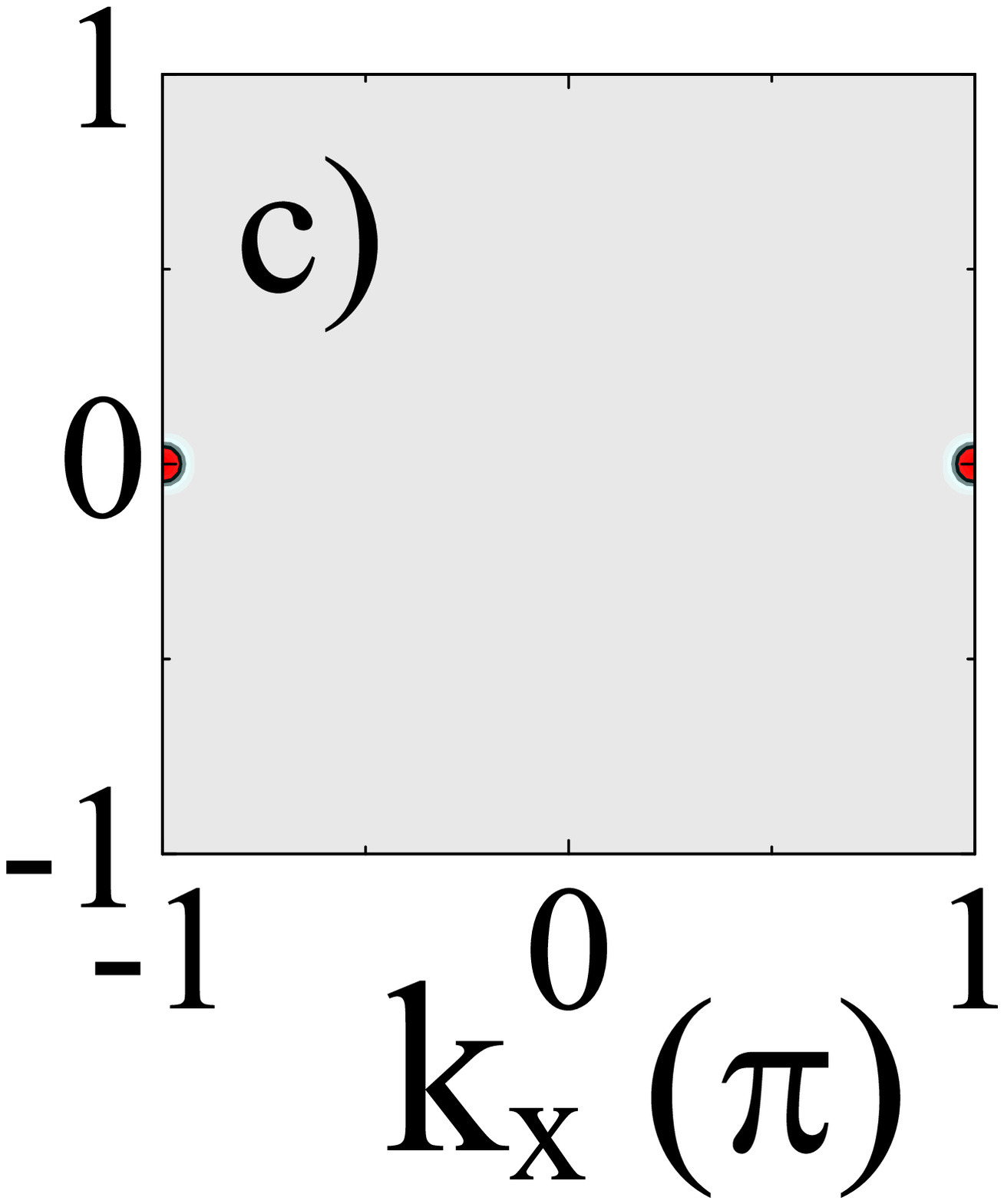}
\includegraphics[width=0.2\linewidth]{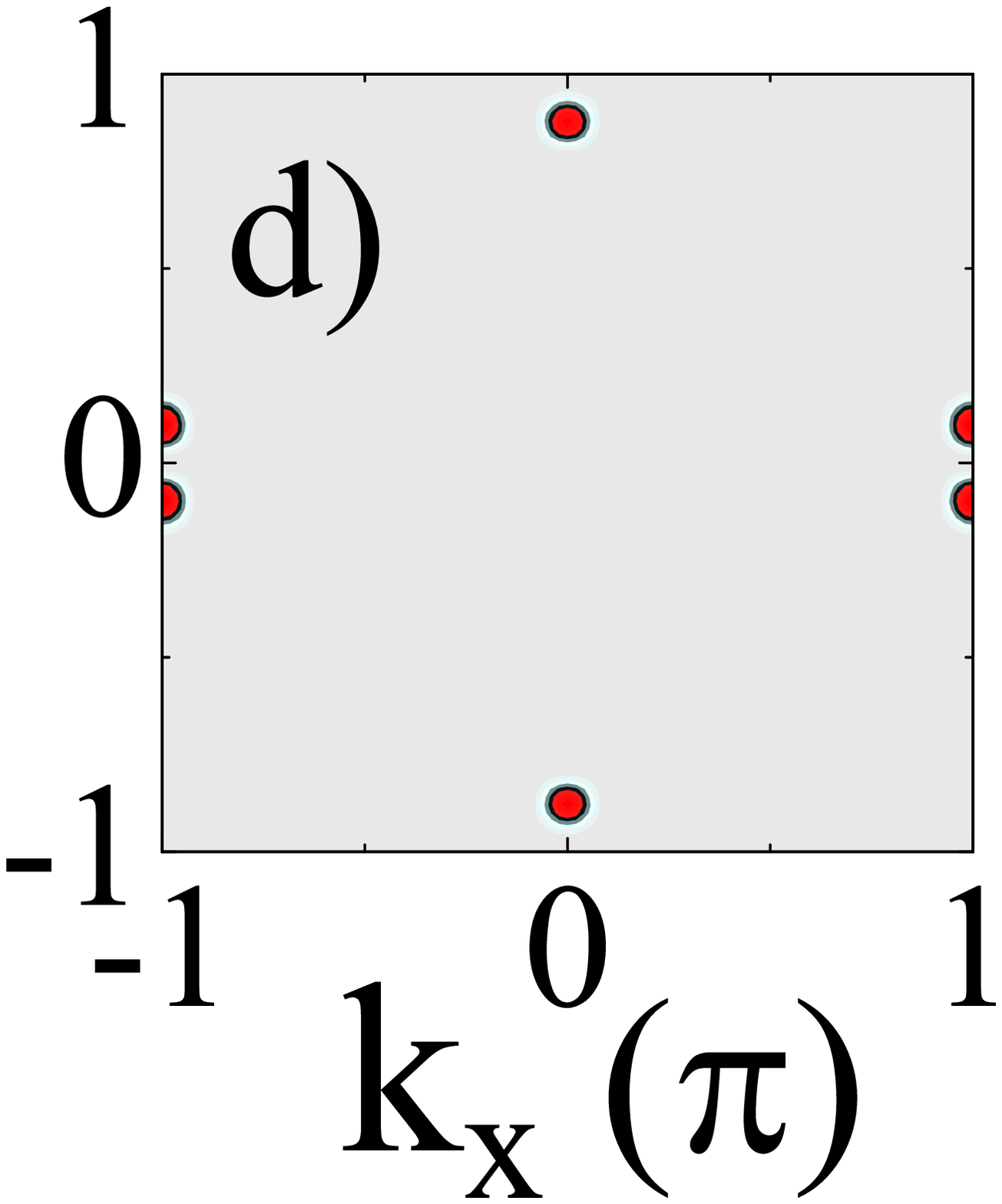}
\caption{(Color online) The spin configurations (distribution of
$\vec{S}_{\bf i}$) in the $xy$-plane with $z=0$: (a) SP
($\theta=0.104$, $h_z=0$); (b) VX ($\theta=1$, $h_z=0$); (c) ST
($\theta=1.24$, $h_z=0$); (d) ST with complicated structure
($\theta=1.56$, $h_z=2.34$), which is not in the phase diagram. (e)
The corresponding spin structure factors $S_{\bf k}^{\bot}$ (see the
text) of the spin configurations in (a-d) are from left to right,
respectively.} \label{fig:spin}
\end{figure}

For simplicity, we assume isotropic parameters in the $xy$ plane,
i.e., $t_x=t_y\equiv t$ and $\theta_x=\theta_y\equiv \theta$, and
take $4t^2/U$ as energy unit hereafter, such that
$J_x=J_y=\cos(2\theta)$, $K_x=K_y=2\sin^2\theta$ and
$D_x=D_y=\sin(2\theta)$. We also assume $t_z=\lambda_1 t$ and
$\theta_z=\lambda_2 \theta$, such that
$J_z=\lambda^2_1\cos(2\lambda_2\theta)$,
$K_z=2\lambda^2_1\sin^2(\lambda_2\theta)$ and
$D_z=\lambda^2_1\sin(2\lambda_2\theta)$. Calculations were mostly
carried out for an $18\times18\times18$ lattice with periodic
boundary conditions. Metropolis MC algorithm \cite {Thijssen} was
used throughout the calculations with $5\times10^6$ (and
$5\times10^5$ for an $xy$-layer with $18\times18$ lattice) sampling
steps at each annealing process with fixed low temperature $T=0.005$
(in units of $4t^2/U$). Some checks on each $xy$ plane with sites
$36\times36$ were performed to ensure consistency. The numerical
results were also confirmed to be nearly the same for the open and
periodic boundary conditions.

\subsubsection{Phase diagram of the reduced 2D layer model with $t_z=0$}

We first consider the case of $t_z\simeq0$ in the Hamiltonian
(\ref{TotHam}), which can be realized by increasing the intensity of
the lasers that generate the periodic lattices along $\hat{z}$ axis
and freeze the atomic motions in this direction. In this case, the
3D system is equivalent to a collection of independent 2D plane
layers along $\hat{z}$ axis, and thus the spin configurations in
each layer are always the same in the ground states. We have
confirmed this point in our numerical simulations. So we can first
look into a single layer and figure out the phase diagram of
Hamiltonian (\ref{TotHam}) with $t_z=0$. In this limit, the spin
Hamiltonian is similar to those in the previous work
\cite{cole2012,radic2012} without the additional Zeeman term. We
have checked that in this case our results are consistent with those
in Refs. \cite{cole2012,radic2012} when the parameter regions are
overlapped, i.e., $h_z=0$, $t_x=t_y$, and $\theta_x=\theta_y$.

We obtain the classical ground state phase diagram of the spin
Hamiltonian $\mathcal{H}^T_s$ with $t_z=0$ as shown in Fig. 1. Apart
from the conventional ferromagnetic (FM) and anti-ferromagnetic
(AFM) phases, there are four unconventional phases having
interesting spin configurations in the $xy$-plane: spiral (SP),
stripe (ST), vortex crystal (VX) and Skyrmion crystal (SKX). Here FM
denotes a ferromagnet with the spins being aligned along one
direction (along the $\hat{z}$-axis in our case due to the Zeeman
field), and the corresponding AFM is an anti-ferromagnet with
neighboring spins pointing in opposite directions (along the
$\pm\hat{z}$-axis in this paper). SP denotes a coplanar ground state
with the spins having a spiral wave in its configuration, whose
spatial periodicity are more than two sites. ST denotes another
coplanar ground state, with spins being separated by periodically
spaced domain walls. The case shown in Fig. 2(c) is a ferromagnetic
stripe phase characterized by a single wave-vector along
$\hat{x}$-axis. VX denotes a crystal state of vortices, with the
spins wind clockwise or counterclockwise around each plaquette in
the $xy$-plane. Finally, SKX is a crystal state of Skyrmions, where
the spins align in a VX shape with nonzero Skyrmion density given by
Eq. (\ref{SKdensity}) below. The presented phase diagram shows a
rich interplay between different magnetic orders, and the parameter
region of the six phases can be found in Fig. 1. We should note that
their boundaries between different phases may be unprecise in the
level of quantum phases, however, this classical approach is
efficient to determine possible phases
\cite{cole2012,radic2012,gong2012}.

The spin textures in the four unconventional magnetic phases have
non-trivial structures and can be characterized by their spin
structure factors $S_{\bf k}^{\perp}\equiv|\sum_{\bf i}\vec{S}_{\bf
i}^{\perp}e^{i\vec{k}\cdot\vec{r}_{\bf i}}|^2$ with $\vec{S}_{\bf
i}^{\perp}=(S_{\bf i}^x,S_{\bf i}^y,0)$. Figures 2(a)-(d) show some
typical spin configurations in the spiral, vortex crystal and stripe
phases, and figure 2(e) shows their spin structure factors $S_{\bf
k}^{\perp}$ in the momentum space (i.e., $k_x$-$k_y$ plane), with
the spots denoting the peaks of the spin structure factors. In Fig.
2(e), $S_{\bf k}^{\perp}$ exhibits a peak at $(0,0.125\pi)$,
corresponding to the spins spiralling along $y$ axis with the wave
number $0.125\pi$ as shown in Fig. 2(a). Similarly, the spins
forming a vortex-crystal configuration in Fig. 2(b) has $S_{\bf
k}^{\perp}$ peaks at $(0.75\pi,0)$ and $(0,0.75\pi)$, and the stripe
spin configuration in Fig. 2(c) has a single peak at $(\pi,0)$,
which means that the spins are staggered along $x$ axis but are
parallel along $y$ axis. We also find that the ground states may
exhibit more complicated stripe spin-textures outside the parameter
region of the phase diagram, with an example being shown in Fig.
2(d).

\begin{figure}[tbph]
\includegraphics[width=0.42\linewidth]{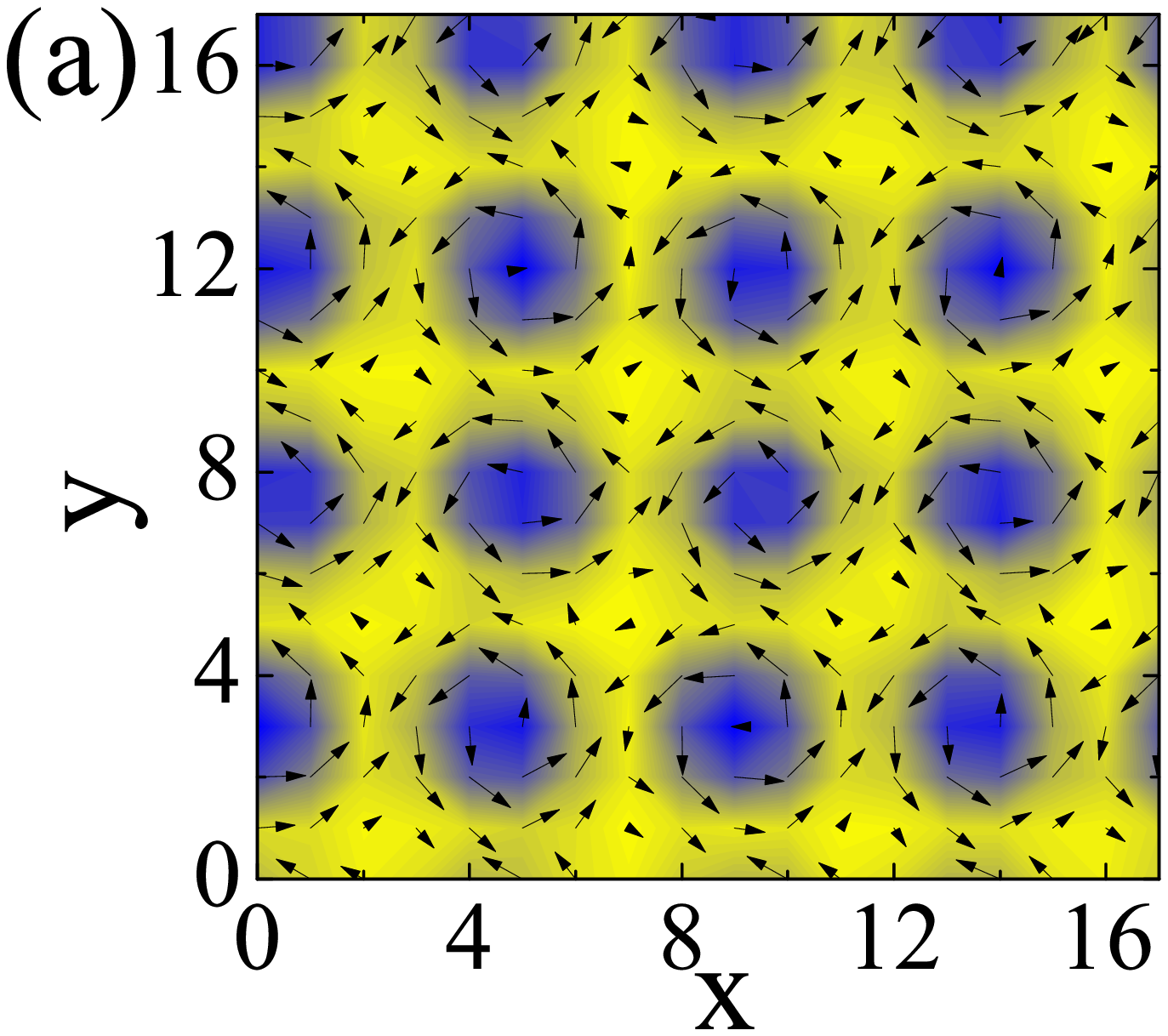}
\includegraphics[width=0.53\linewidth]{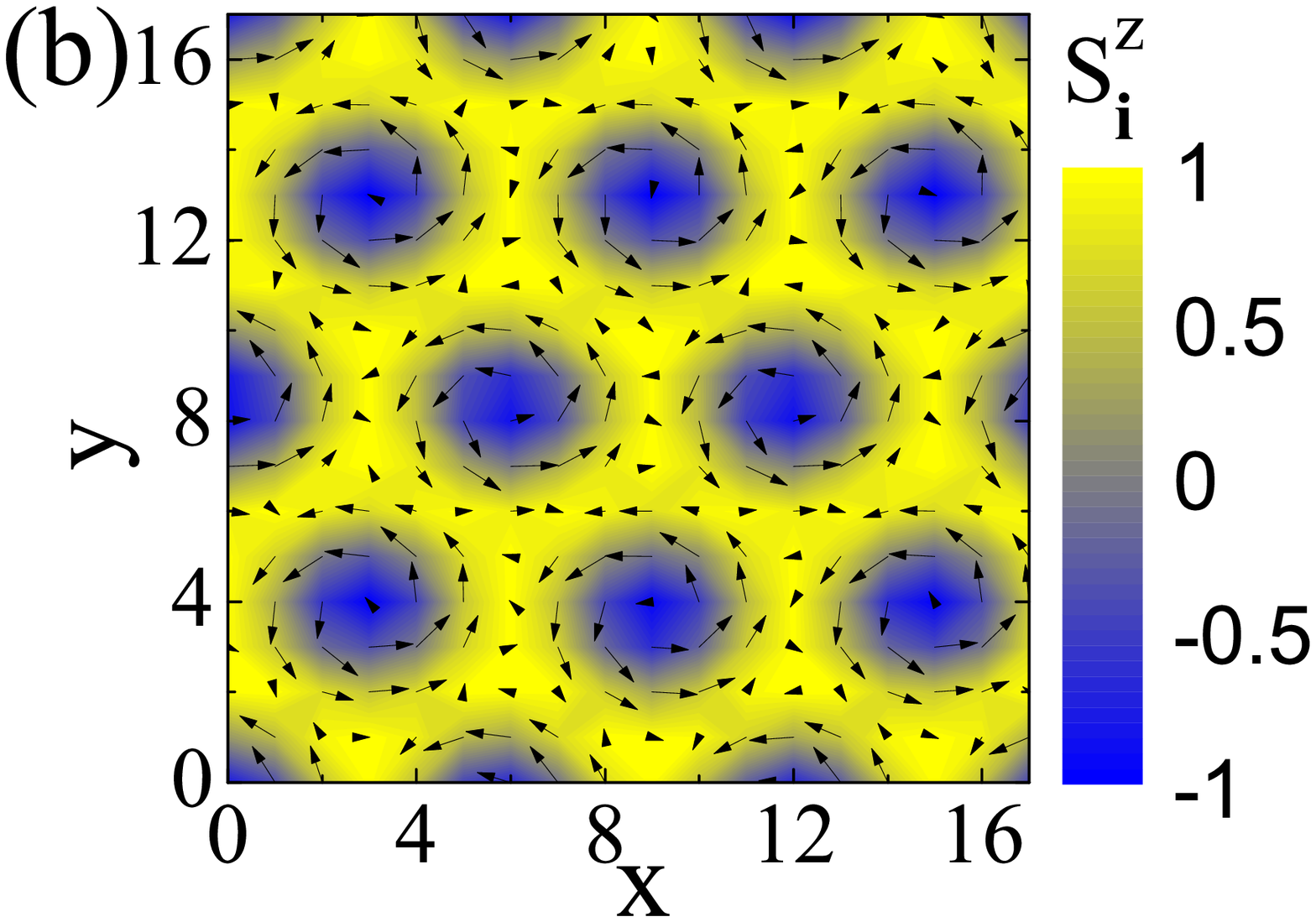}
\includegraphics[width=0.4\linewidth]{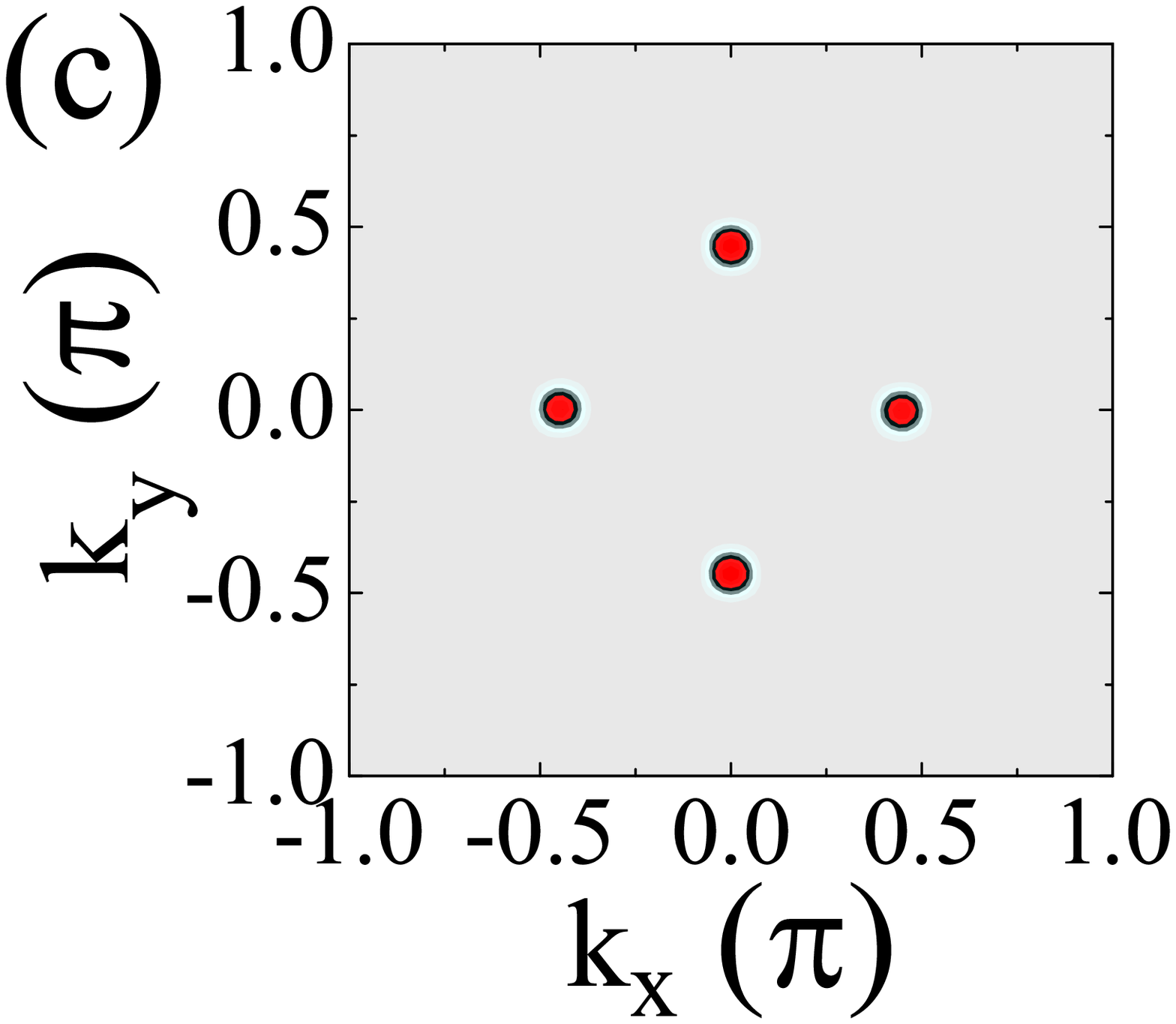}~~
\includegraphics[width=0.4\linewidth]{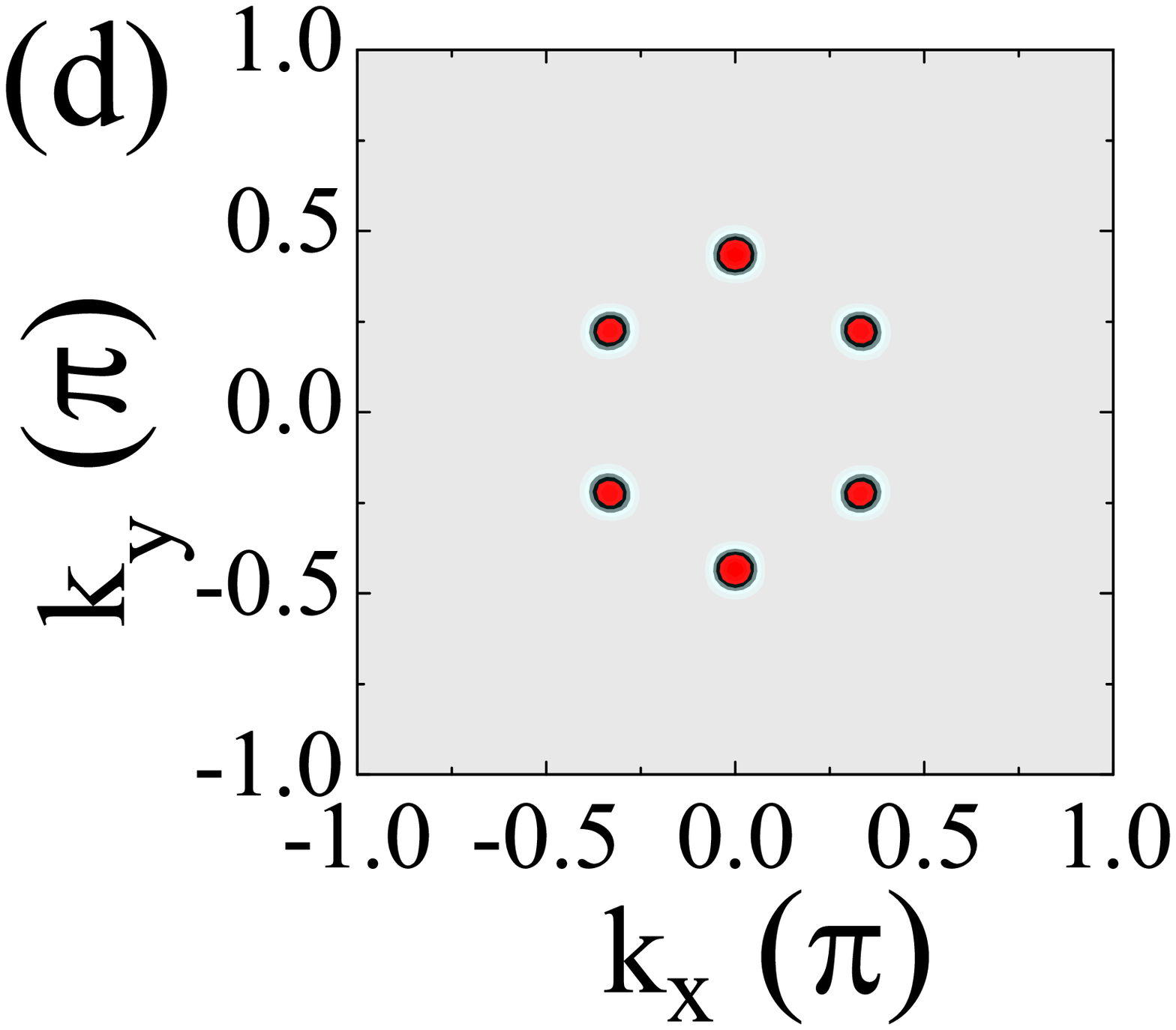}
\includegraphics[width=0.48\linewidth]{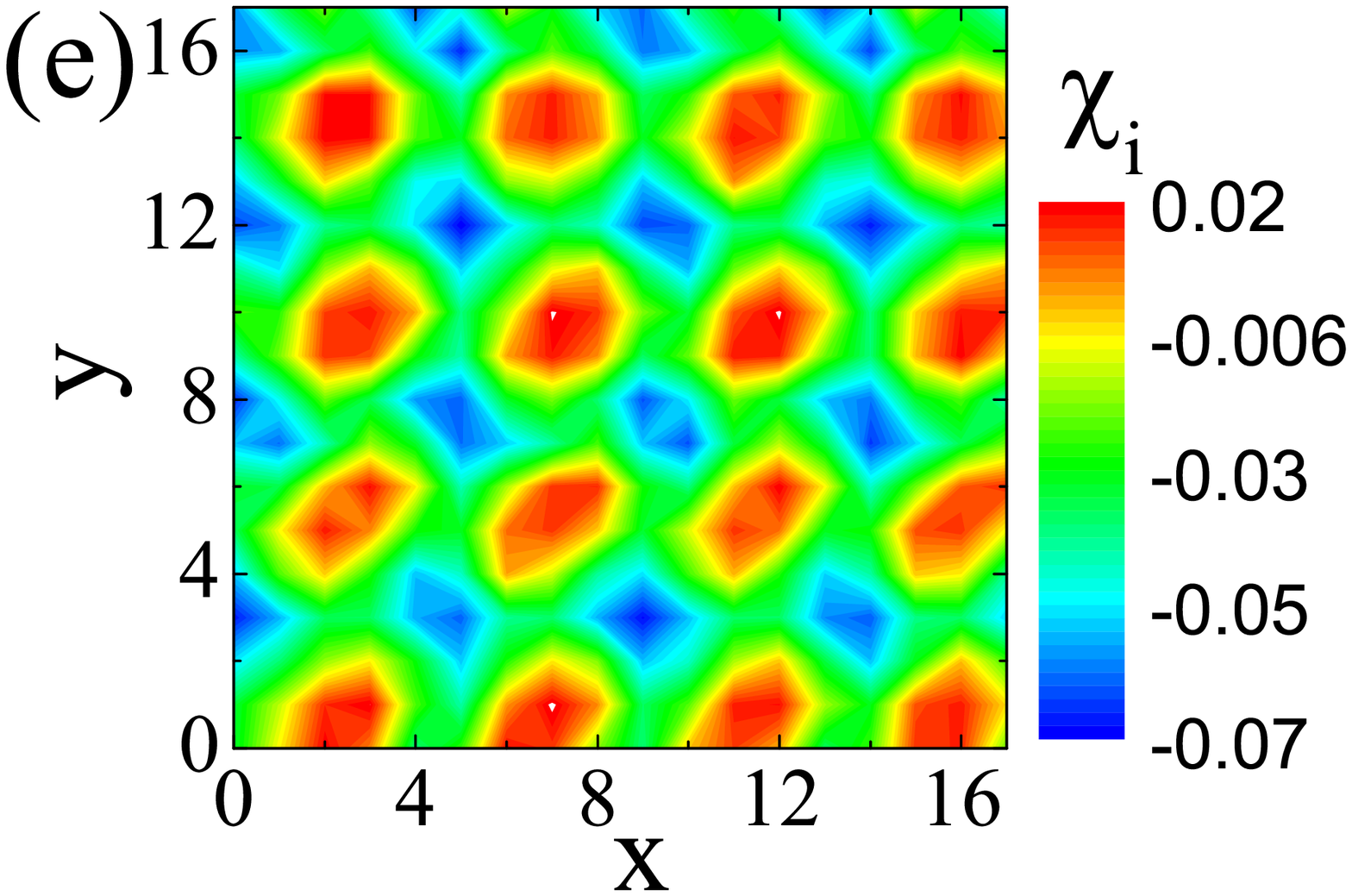}
\includegraphics[width=0.48\linewidth]{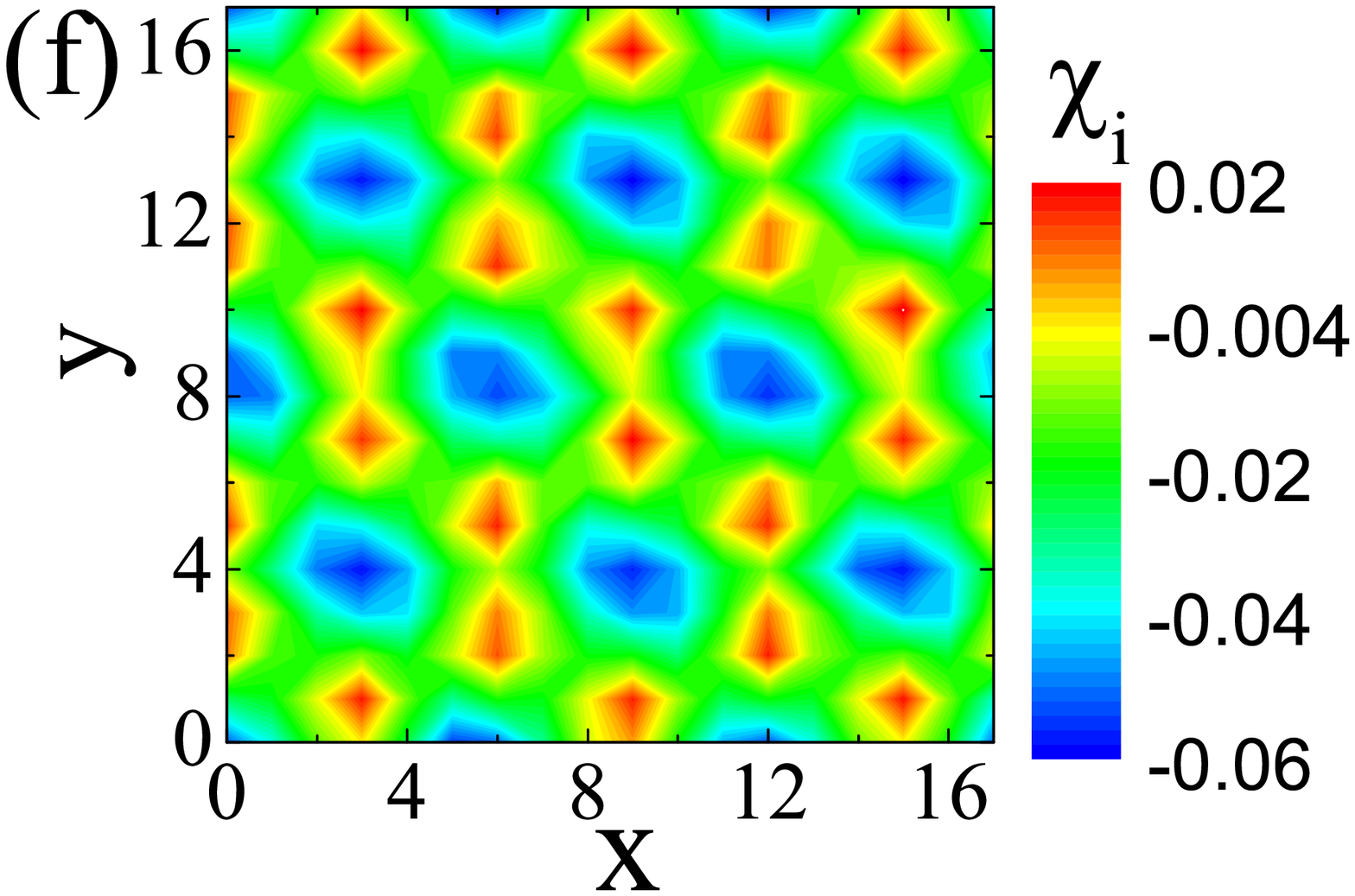}
\caption{(Color online) Properties of the 2D Skyrmion crystals.
Typical spin configurations in the Skyrmion crystal phase with (a)
square and (b) hexagonal symmetry, and the color scale shows the
magnitude of out-plane component $S_z$. (c) and (d) are the
corresponding spin structure factors $S_{\bf k}^{\perp}$ of the spin
configurations in (a) and (d), respectively. (e) and (f) are the
local density $\chi_{{\textbf i}}$ (see the text) of the square and
the hexagonal Skyrmion crystals, respectively. The parameters are
$\theta=0.726$ and $h_z=0.910$ for the square Skyrmion crystals;
$\theta=0.592$ and $h_z=0.756$ for the hexagonal Skyrmion crystals.}
\label{fig:SKX}
\end{figure}

The Skyrmion crystal phase in the phase diagram, referring to the
phase where the spins presenting an array of 2D Skyrmions (see Fig.
3), has a large parameter region when varying the SO-coupling
strength $\theta$ and the Zeeman field $h_z$. The array of Skyrmions
is anisotropy in general, and can present square or
hexagonal-symmetric patterns 
for proper SO-coupling strengths. Figures 3(a) and (b) show the
typical spin configurations in the Skyrmion crystal phase with
square and hexagonal structures, respectively. The corresponding
spin structure factors $S_{\bf k}^{\perp}$ shown in Figs. 3(c) and
(d) also reflect their symmetries in the momentum space. We note
that the hexagonal Skyrmion crystal state in fact exhibits a
triangular lattice structure of Skyrmions in the real space since
$S_{\bf i}^z$ is taken in account [see Fig. 3(b)]. However, we still
refer to it as hexagonal one in order to be consistent with its
hexagonal Bragg pattern in the Fourier analysis [see Fig. 3(d)].

Before ending this part, we analyze the energies of the obtained
spin states. For the reduced 2D system here, the energy functional
$E[\vec{S}_{\bf i}]$ is given by the Hamiltonian (\ref{TotHam}) with
$J_{x,y}=\cos(2\theta)$, $K_{x,y}=2\sin^2\theta$,
$D_{x,y}=\sin(2\theta)$, and $J_z=D_z=K_z=0$. A classic spin state
can be parameterized by $\vec{S}_{\bf i}=S(\cos\gamma_{\bf
i}\sin\varphi_{\bf i},\sin\gamma_{\bf i}\sin\varphi_{\bf
i},\cos\varphi_{\bf i})$ with $S^2=1$ and ${\bf i}\doteq(x_i,y_i)$.
For the $\hat{z}$-axis ferromagnetic and antiferromagnetic spin
states in our case, the energy densities defined as the energy per
spin are $E_{\text{FM}}=-E_{\text{AFM}}=-h_z-2\cos(2\theta)$ (here
$h_z>0$ and $0<\theta<\pi/2$). For the spiral and stripe spin states
shown in Fig. 2(a) and (c), based on their spin structures in the
momentum space, one can write down their spin configurations as
$\vec{S}_{\bf
i,SP}=(\sin[k_{\text{sp}}y_i],0,\cos[k_{\text{sp}}y_i])$ with
$k_{\text{sp}}=0.125\pi$, and $\vec{S}_{\bf
i,ST}=[(-1)^{x_i+y_i}\cos(k_{\text{st}}x_i+\phi),\sin(k_{\text{st}}x_i+\phi),0]$
with $k_{\text{st}}=\pi$ and $\phi=0.44\pi$ being a phase angle.
Here we have dropped an irrelevant overall phase. Direct
calculations yield the energy densities of the two spin states
$E_{\text{SP}}=-\cos(2\theta)(1+\cos k_{\text{sp}})-\sin(2\theta)
\sin k_{\text{sp}}-\sin^2\theta$, and $E_{\text{ST}}=
-2\sin^2\theta$. Here $E_{\text{ST}}$ is independent of $\phi$,
which indicates that the stripe spin states in our system with
different phase angles are degenerate on the classical level. For
the crystal states of vortices and Skyrmions shown in Figs. 2(b) and
3(a,b), which contain more than one spiral wave, it is very hard to
give an expression of their spin configurations \cite{yi2009}. For
example, one has to replace the local constraint $\vec{S}_{\bf
i}^2=1$ followed in our numerical simulations and previous analysis
by a global one (i.e. $\langle\vec{S}_{\bf i}^2\rangle=1$) in
approximately writing down the spin configuration of the square
Skyrmion crystal as $\vec{S}_{\bf
i,S-SKX}\propto(\sin[k_{\text{skx}}y_i],\cos[k_{\text{skx}}x_i],\sin[k_{\text{skx}}x_i]+\cos[k_{\text{skx}}y_i]$)
\cite{yi2009}, with the wave number $k_{\text{skx}}=0.5\pi$ as shown
in Fig. 3(c). However, this approach will lead to a considerable
deviation in estimating the energy density of these spin states
\cite{yi2009}. Therefore, we numerically compute their energy
densities by directly using the obtained spin configurations in
Figs. 2(b) and 3(a,b). For the parameters $\theta=0.726$ and
$h_z=0.91$, we get the energy density of the vortex crystal state
$E_{\text{VX}}=-1.18$, the square Skyrmion crystal state
$E_{\text{S-SKX}}=-1.77$, and the hexagonal Skyrmion crystal state
$E_{\text{H-SKX}}=-1.73$. One can check that $E_{\text{S-SKX}}$ is
smallest among all the six spin states for the parameters, which
indicates that the square Skyrmion crystal state is indeed the
classical ground state in this case. For $\theta=0.592$ and
$h_z=0.756$, we get $E_{\text{VX}}=-0.99$, $E_{\text{S-SKX}}=-1.82$,
and $E_{\text{H-SKX}}=-1.86$, so the hexagonal Skyrmion crystal
state is the classical ground state in this case.

\subsubsection{2D Skyrmion crystals and effects of Zeeman fields}

The 2D Skyrmion crystals obtained in this model are distinguished
from those in Ref. \cite{cole2012} not only in their spin
configurations but also in the mechanism. The 2D model in Ref.
\cite{cole2012} contains no Zeeman field, however, the Zeeman field
is crucial here. As seen from the phase diagram, there are no
Skyrmion crystal states without the Zeeman field; they should be
generated  from the spiral waves with the help of a Zeeman field,
while the spiral waves are formed via the competition between the DM
interactions and the Heisenberg exchange interactions. In the
presence of isotropic DM interactions ($D_{\eta}$ is independent of
$\eta$), the spiral spins can give way to the Skyrmion crystal spins
in energy for a sufficiently strong Zeeman field or anisotropic spin
interactions \cite{yi2009,Park2011}. Because all the spin
interactions in our system are interdependent with respect to the SO
coupling strengths [for example, $K_{\eta}$ is larger than
$D_{\eta}$ when $\theta_{\eta}>\pi/4$ in the Hamiltonian
(\ref{3DSHam})], the Skyrmion crystal states can be stabilized only
by the Zeeman field (see Fig. 1). The presence of Skyrmion crystal
spins in Ref. \cite{cole2012} without a Zeeman field is due to the
anisotropic DM interactions, which can increase the effective
anisotropic spin interactions. With increasing of Zeeman energy in
the SKX regime, some Skyrmion crystal spins begin to melt into the
ferromagnetic spins when the Zeeman term becomes dominant over the
DM term, and finally all of them realign as a ferromagnet along the
$\hat{z}$-axis over a critical value, which is not shown in the
phase diagram.

Such kind of Skyrmion crystals have been explored by numerical
calculations and in experiments in chiral magnet materials
\cite{yi2009,roler2006,yu2010}, but the controllability in the
materials \cite{roler2006,yu2010} is low in contrast to that in the
cold atom system. This system is clean and the paraments, such as
$\theta$ and $h_z$, are widely tunable via adjusting laser-atom
interations \cite{Jaksch1998,Greiner2002}. So the density of
Skyrmions and the symmetry of the Skyrmion crystal can be well
controlled by varying the two parameters. For instance, increasing
the Zeeman field can lead to the increase of the Skyrmion density up
to certain levels, and varying the SO-coupling strength can change
the distribution of Skyrmions, from generally anisotropy to square
or hexagonal symmetry as shown in Figs. 3(a) and (b).

To further characterize the Skyrmion crystals, we introduce the
local density of Skyrmions $\chi_{{\textbf i}}$ at lattice site
${\bf i}$ in each $xy$-plane as \cite{yi2009,alba2011}
\begin{equation}
\label{SKdensity} \chi_{{\textbf
i}}=\frac{1}{8\pi}\left[\vec{S}_{\bf i}\cdot(\vec{S}_{{\bf
i}+\hat{x}}\times\vec{S}_{{\bf i}+\hat{y}})+ \vec{S}_{\bf
i}\cdot(\vec{S}_{{\bf i}-\hat{x}}\times\vec{S}_{{\bf
i}-\hat{y}})\right],
\end{equation}
which is the discretization counterpart of the well-known
topological charge density
$\vec{S}\cdot(\partial_x\vec{S}\times\partial_y\vec{S})/4\pi$ for
the continuum case \cite{kawakami2012}. For a single localized 2D
Skyrmion here, its topological winding number given by $W_{{\text
2D}}=\sum_{\bf i}^{\text{unit cell}}\chi_{{\textbf i}}$ plus the
sign of its pole (i.e., here $S^z_{\bf i}=-1$ at the Skyrmion cone
as shown in Fig. 3) equals to unit in the continuum limit. Note that
for an ordinary vortex, this topological number equals to zero. The
winding number is stable with respect to the discretization, as for
a lattice layer with size $L\times L$, the fluctuation (error) is on
the order of $\mathcal{O}(4\pi^2/L^2)$ \cite{alba2011}. In Figs.
3(e) and (f), we show local density of the square and the hexagonal
Skyrmion crystals. The winding number of a single Skyrmion in the
two cases is numerically computed to be nearly minus one.

\begin{figure}[tbph]
\includegraphics[width=1.0\linewidth]{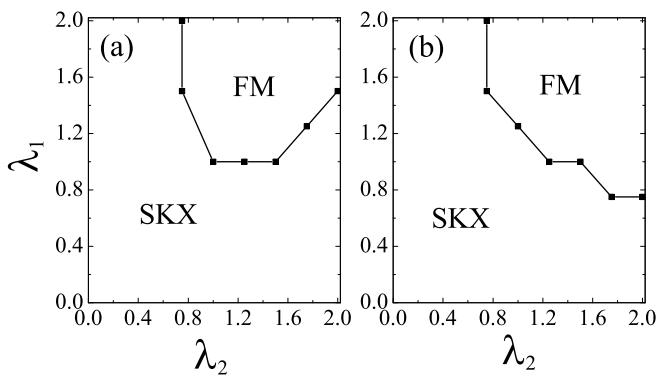}
\caption{Stability of layer Skyrmion crystals with respect to the
spin-interaction parameters $\lambda_1$ and $\lambda_2$ along the
$\hat{z}$ axis. The parameter regions for (a) square and (b)
hexagonal Skyrmion crystals denoted by SKX are both wide. Another
region denoted by FM is the ferromagnetic phase. The parameters are
$\theta=0.726$ and $h_z=0.910$ in (a); $\theta=0.592$ and
$h_z=0.756$ in (b).} \label{fig:phase}
\end{figure}

\subsubsection{Inter-layer spin interactions and why no 3D Skyrmions}

Finally in this section, we turn to consider the inter-layer
spin-spin interactions along the $\hat{z}$ axis and check the
stability of the square and hexagonal Skyrmion crystals obtained
previously [see Fig. 3(a) and (b)] with respect to the parameters
$\lambda_1$ and $\lambda_2$. Our numerical results are shown in Fig.
4. We find that the Skyrmion crystals with both square and hexagonal
structures are stable and have a large region in the parameter space
($\lambda_1$,$\lambda_2$). Since the parameters $\lambda_1$ and
$\lambda_2$ are position-independent, the spin configurations still
exhibit the same distribution of Skyrmions in each $xy$ layer, which
has been confirmed in our numerical calculations. That is to say,
the 2D Skyrmion crystal states extend along the $\hat{z}$ axis in a
columnar manner \cite{Park2011}. In the parameter space (see Fig.
4), there is another phase, i.e. the ferromagnetic phase. This
demonstrates that the layer Skyrmion crystals can be melt into the
conventional ferromagnet by spin-spin interactions along the
$\hat{z}$ axis in Hamiltonian (\ref{TotHam}). We note that near the
boundary between the two phases, the Skyrmions and ferromagnet
coexist in the spin configurations and hence the Skyrmion crystals
are ill-defined there. However, in most of the region denoted by SKX
in Figs. 4(a) and (b), the square and hexagonal Skyrmion crystals
remain.

In our numerical calculations done by the classical MC method, we do
not find a topologically non-trivial spin structure forming a
genuine 3D Skyrmion or its corresponding crystal state \cite
{Klebanov} for the spin Hamiltonian (\ref{TotHam}) in a large
parameter region. For the most isotropic cases with $\theta_{\eta} =
\theta =\pi/4$ and $h_z = 0$ (one has $D_{\eta} = K_{\eta} =1$ and
$J_{\eta} = 0$), the classical ground state is a spiral state with
four-site periodicity along one axis ($\hat{x}$, $\hat{y}$, or
$\hat{z}$ axis). An additional small Zeeman field favors a certain
direction. Note that another degenerate spiral state with six-site
periodicity in [111] direction do not appear in the numerical
simulations, probably owing to the finite size effects. For other
values of $\theta$ with $h_z = 0$, we find similar behavior for the
ground states, i.e., the spins in each plane aligning in the same
pattern and extending along the perpendicular axis in a columnar
manner. At a strong Zeeman field, all these spins states would be
melt into a ferromagnetic state along the direction of the Zeeman
field. We have also checked that the special Zeeman field along
[111] direction has similar effects as those along other directions,
such as modulating the polarization of the spiral spins and changing
the density of the 2D Skymions, without bringing the 3D Skyrmions.

Similar 3D spin models in the context of chiral magnetism have been
investigated by variational energy analysis within the global-spin
constraint condition \cite{yi2009,Park2011}. Detailed calculations
in Ref. \cite{yi2009} indicate that a 3D Skyrmionic ground state is
possible; however, there would be other 3D spin states with lower
energy as shown in Ref. \cite{Park2011}. So, whether 3D Skyrmion
crystal states can be energetically stable is yet to be explored in
spin models \cite{Rybakov}. Back to our numerical results, there is
no indication of such a classical state at low temperature. This is
consistent with a very recent numerical work by the classical MC
simulation for a similar 3D lattice spin model \cite{Buhrandt}. The
discrepancy might also be due to the soft spins constraint
$\langle\vec{S}_{\bf i}^2\rangle=1$ used in the variational
calculations in Refs. \cite{yi2009,Park2011}, which replaces the
hard spins constraint $\vec{S}_{\bf i}^2=1$ in the MC simulation. We
note that this cold atom system with tunable parameters may provide
a better platform for exploring the Skyrmion physics and searching
for 3D Skyrmion crystal states in a semi-classical (such as using
the soft spins) or quantum approach, which would be an interesting
challenge in our further studies.

\section{discussion and conclusion}

Before concluding this paper, we briefly discuss some methods for detecting the
superfluidity states and the spin configurations in the deep MI phase. The plane-wave
and the strip superfluidity phases correspond to BECs at a single finite momentum
and at a pair of opposite momenta, respectively. The values of the momenta depend on
the SO-coupling strength as discussed in Sec. III. The standard time-of-flight imaging
measurement can reveal the condensate peaks at the nontrivial momentum points \cite{Greiner2002},
which provides direct signatures of the two superfluidity states. The different magnetic
orders presented above can be detected by the optical Bragg scattering for atoms in OLs \cite {Hulet},
as the peaks in the Bragg spectroscopy directly reveal their spin structure factors.
Another way to measure the spin configurations is using the spin-resolved {\sl in-situ} imaging technique \cite{Bakr}.
The idea is to implement a high-resolution optical imaging system, by which single atoms are
detected with near-unity fidelity on individual sites of an OL \cite{Bakr}. A similar measurement
has been performed for revealing phase transitions of an atomic Ising chain \cite{Simon}.

In summary, we have studied the superfluidity and magnetic
properties of ultracold bosons with the synthetic 3D SO coupling in
a cubic OL. The lowest energy band exhibits one, two or four pairs
of degenerate single-particle ground states depending on the
SO-coupling strengths, which can lead to the condensate states with
spin-stripes for the weak atomic interactions. In the MI state with
one particle per site, an effective spin Hamiltonian with the 3D DM
spin interactions is derived. The spin Hamiltonian with an
additional Zeeman field has a rich phase diagram with spiral,
stripe, vortex crystal, and Skyrmion crystal spin-textures in each
$xy$-plane layer. The Skyrmion crystals extended along the $z$ axis
in a columnar manner can be tunable with square and hexagonal
symmetries, and stable against the inter-layer spin-spin
interactions in a large parameter region.

~~

\section{acknowledgments}
We thank Tao Zhou and Yan Chen for helpful discussions. This work
was supported by the NSFC (Grant No. 11125417), the SKPBR (Grant
No. 2011CB922104), the PCSIRT, the SRFGS of SCNU, the GRF
(HKU7058/11P),  CRF (HKU-8/11G) of the RGC of Hong Kong, and the URC fund of HKU.

\end{document}